\RequirePackage{fix-cm}
\documentclass[smallextended]{svjour3}       
\smartqed  
\usepackage{graphicx}
\usepackage{todonotes}
\usepackage[hyphens]{url}
\usepackage{hyperref}
\setuptodonotes{inline}

\newtheorem{innercustomthm}{Hypothesis}
\newenvironment{hypothesis}[1]
  {\renewcommand\theinnercustomthm{#1:}\innercustomthm}
  {\endinnercustomthm}

\usepackage{amsmath}
\usepackage{url}
\usepackage{hyperref}
\usepackage{enumitem}
\usepackage{multirow}
\usepackage{booktabs}
\usepackage{array}
\newcommand{\spanval}[1]{\dotfill\,#1\,\dotfill}
\usepackage[sort&compress]{natbib}
\setcitestyle{authoryear}

\begin{document}

\title{The Influence of Code Comments on the Perceived Helpfulness of Stack Overflow Posts}

\titlerunning{Influence of Code Comments on the Perceived Helpfulness of SO Posts}

\author{Kathrin Figl \and Maria Kirchner \and Sebastian Baltes \and Michael Felderer}


\institute{ Kathrin Figl \at University of Innsbruck, Austria \\
                \email{kathrin.figl@uibk.ac.at} \\
                ORCID: 0000-0002-0777-2295
            \and
            Maria Kirchner \at University of Innsbruck, Austria \\
                \email{kirchnermaria7@gmail.com} \\
                ORCID: 0009-0005-3608-9144
            \and
            Sebastian Baltes \at University of Bayreuth, Germany \\
                \email{sebastian.baltes@uni-bayreuth.de} \\
                ORCID: 0000-0002-2442-7522
            \and
            Michael Felderer \at Germany Aerospace Center (DLR), Germany \\ 
                \email{michael.felderer@dlr.de}
                            \at University of Cologne, Germany \\
                \email{michael.felderer@uni-koeln.de} \\
                ORCID: 0000-0003-3818-4442
}

\makeatletter
\def\makeheadbox{%
  \hbox to0pt{%
    \vbox{\baselineskip=10dd
      \hrule
      \hbox to\hsize{%
        \vrule\kern3pt
        \vbox{\kern3pt
          \hbox{\bfseries Empirical Software Engineering}%
          \hbox{(accepted manuscript)}%
          \kern3pt}%
        \hfil\kern3pt\vrule}%
      \hrule}%
    \hss}}
\makeatother

\date{Received: 18 Aug 2024 / Accepted: 26 Aug 2025}

\maketitle

\begin{abstract}
Question-and-answer platforms such as Stack Overflow are an important way for software developers to share and retrieve knowledge.
However, reusing poorly understood code can lead to serious problems, such as bugs or security vulnerabilities.
To better understand how code comments affect the perceived helpfulness of Stack Overflow answers, we conducted an online experiment simulating a Stack Overflow environment (n=91).
The results indicate that both block and inline comments are perceived as significantly more helpful than uncommented source code.
Moreover, novices rated code snippets with block comments as more helpful than those with inline comments.
Interestingly, other surface features, such as the position of an answer and its answer score, were considered less important.
Moreover, the content of Stack Overflow has been a major source for training large language models.
AI-based coding assistants such as GitHub Copilot, which are based on these models, are changing the way Stack Overflow is used.
However, our findings have implications beyond Stack Overflow.
First, they may help to improve the relevance also of other community-driven platforms, which provide human advice and explanations of code solutions, complementing AI-based support for software developers.
Second, since chat-based AI tools can be prompted to generate code in different ways, knowing which properties influence perceived helpfulness can lead to more targeted prompting strategies to generate readable code snippets.
\end{abstract}

\section{Introduction}

Programming is a constantly evolving process of problem solving that encompasses tasks such as feature implementation, debugging, and system maintenance.
In this dynamic field, software developers often turn to a variety of information sources to achieve their goals.
The advent of artificial intelligence (AI) has caused a paradigm shift in programming practices.
AI-powered tools have the ability to generate code snippets from user input, streamlining the prototyping process for developers, and scan code for inefficiencies and bugs, while predictive coding and auto-completion features suggest code segments.
However, the continued relevance of community-driven platforms such as Stack Overflow, even in the face of technological advances, is underscored by a modest 2.6\%~\citep{RN14011} decline in question activity on the site since the introduction of ChatGPT.
It highlights the importance of platforms that provide human advice on programming issues.
Stack Overflow, with its large pool of experienced developers, offers an environment in which detailed, community-validated, and context-specific advice thrives.
The platform's peer-review system, manifested through user upvotes and downvotes, is built to ensure that the most accurate and helpful answers are elevated, maintaining a standard of quality and reliability. 
The diversity of answers to a single question, including their comments---each providing a unique perspective or solution---often yields more comprehensive insights than the solutions typically generated by AI.
However, it is also worth noting that there is an active discussion in the community on if and why Stack Overflow is ``fading away''\footnote{\url{https://news.ycombinator.com/item?id=41364798}} and whether LLMs make Stack Overflow irrelevant.\footnote{See also: \url{https://blog.pragmaticengineer.com/are-llms-making-stackoverflow-irrelevant/}; \url{https://meta.stackoverflow.com/q/422392/1974143}}

Despite these discussions, code examples are a critical resource that eases the challenges of independent problem solving.
Stack Overflow, with its extensive repository of more than 24 million questions, 36 million answers, and 91 million comments~\citep{RN14008}, remains the main question-and-answer (Q\&A) forum for software developers around the world.
As a platform for learning and sharing, Stack Overflow has revolutionized the way developers seek help and provide solutions to programming questions.
It acts as a knowledge-sharing hub, making it easier for developers to share their expertise and find answers to their problems.
By sharing their personal experiences and anecdotes, contributors to the platform offer practical insights into the application of programming knowledge that is more personal than AI-generated advice.
The site serves as a comprehensive resource for developers seeking help beyond their local communities.
Stack Overflow users include those who ask questions, those who answer them, and those who browse existing Q\&A posts for help. 

Stack Overflow has attracted significant research interest from both the information systems community~\citep{RN14012, RN14014, RN14010, RN14013} and the empirical software engineering community.
Many of the previous studies focused on different properties of the content on Stack Overflow such as text, code, and metadata~\citep{DBLP:conf/icse/TreudeBS11, DBLP:conf/msr/BaltesDT008, DBLP:journals/ese/WuWBI19} and its relationship to other platforms such as GitHub~\citep{DBLP:conf/socialcom/VasilescuFS13, DBLP:journals/ese/BaltesD19, RN13874, RN13870}. 
However, there is a lack of research that examines factors from the \textbf{user's perspective} when selecting a code snippet.
Answer selection is critical because even if developers find a relevant post that addresses their problem, they still need to choose an answer that includes a code snippet that can solve their specific problem.
Prior research, including that of \cite{RN14002}, has identified poor comprehensibility of Stack Overflow code snippets as a common problem that complicates their reuse. Potential risks include copying poorly understood code, which can lead to bugs~\citep{RN13870}, security vulnerabilities~\citep{RN14003}, and licensing issues~\citep{DBLP:journals/ese/BaltesD19}.

To address this research gap, this study investigates the factors that Stack Overflow users consider when selecting code snippets and solutions to their programming problems. Specifically, it examines the role of code style features, such as the type of source code comments (block, inline, or none) and surface-level cues, including answer position and score.
Improving the understanding of code snippets with comments is particularly relevant as users increasingly rely on Stack Overflow for complex questions, while simpler questions are often handled by AI tools~\citep{RN14011}.
The following research questions guided the study:

\begin{description}[style=multiline, leftmargin=10mm]
\item[\textbf{RQ1}] What is the impact of code comments on the perceived helpfulness of code snippets on Stack Overflow?
\item[\textbf{RQ2}] Does the impact of code comments on perceived helpfulness differ between novices and experts?
\item[\textbf{RQ3}] Does the position of a Stack Overflow answer within a Stack Overflow thread affect perceived helpfulness?
\item[\textbf{RQ4}] Does the presence of answer scores amplify the effect of position on perceived helpfulness?
\end{description}

To investigate these questions, we conducted an online experiment with 91 participants in a simulated Stack Overflow environment.
Participants rated code snippets that varied in comment type, answer position, and score. We also examined whether programming experience (novice vs. expert) moderates the perceived helpfulness of different types of comments and surface features.
The results show that code snippets with comments were generally perceived as more helpful than those without comments, with block comments receiving the highest helpfulness ratings.
This effect was particularly pronounced among novices, who rated block comments as significantly more helpful than inline comments. In contrast, surface-level cues such as answer position and answer score had no significant effect on perceived helpfulness. These findings contribute to a better understanding of how developers evaluate shared code snippets and suggest that code comments---especially block comments---play a crucial role in enhancing the perceived helpfulness of programming answers on community-driven platforms.

\section{Related work and theoretical background}
\label{sec:background}

In this section, we summarize related work on Stack Overflow as a social Q\&A platform for software developers, as well as theories and empirical studies on code reading and comprehension.

\subsection{Stack Overflow as a social Q\&A platform for software developers}
\label{sec:stack-overflow}

Social media has changed the way software developers communicate online and coordinate software development activities~\citep{DBLP:journals/tse/StoreyZFSG17}.
Social communities for software developers, such as GitHub, Twitter/X, or Stack Overflow, provide participants with quick access to knowledge and expertise shared by other developers.
These platforms enable various activities, including searching for projects hosted by other developers, contributing to projects, hosting one's own projects, participating in discussions about programming technologies, and following repositories of interest.
Together, individuals form a community around their shared interests on social developer platforms~\citep{DBLP:conf/socialcom/VasilescuFS13}.
Technical Q\&A platforms such as Stack Overflow have become increasingly important to software developers as a means of knowledge sharing.
As one of the most popular Q\&A sites in the world, Stack Overflow serves as a community-based platform for collaborative information sharing spanning diverse programming languages.
Over the years, Stack Overflow has accumulated a vast amount of programming knowledge consisting of code snippets accompanied by natural language explanations.
The popularity and significance of Q\&A forums as a source of support for software developers to solve their development problems are substantial.
The primary function of Stack Overflow is to provide a platform where users can ask questions and receive answers from other users.
To improve the quality of both questions and answers, users have the ability to comment on and edit each other's posts.
In addition, users can earn reputation points and badges for their contributions, e.g., if other users upvote their content. 

Researchers have conducted studies on questions, answers, and code examples on Stack Overflow to examine various aspects related to the quality of answers, including the types of questions and comments~\citep{DBLP:conf/icse/TreudeBS11, RN13992, RN13874}, their helpfulness~\citep{RN13992}, and the characteristics of good and bad question-answer relationships~\citep{RN13979, RN13876}.
In addition, research has been conducted on coding style and code fragments~\citep{RN13974, RN13873, RN13986, RN14002}, as well as developer and user behavior observed on Stack Overflow~\citep{RN14012, RN14010, RN14013, RN14003}.
Furthermore, researchers have applied topic analysis and mining techniques to study the reuse of code snippets~\citep{RN13870, DBLP:journals/ese/BaltesD19, DBLP:journals/ese/WuWBI19}.
Moreover, studies have analyzed the characteristics of good and bad quality code snippets~\citep{RN13979, RN13876}.
\cite{RN13979} examined code snippets and the code-to-text ratio on Stack Overflow, analyzing metrics to assess their impact on quality. 
Other studies have revealed that Stack Overflow content has been a major source for training large language models~\citep{nasr2023scalable}.

A fundamental issue that stands out in the literature is that code snippets on Stack Overflow can be difficult to understand, making it challenging for developers to reuse parts of the code, potentially leading to security issues.
\cite{RN14002} conducted a study to investigate the extent to which developers perceive Stack Overflow code snippets as self-explanatory, identifying several issues such as the organization of the code, naming, incompleteness, quality, domain, clutter, and rationale.
\cite{RN14003} researched how developers use Stack Overflow as a resource and their interactions with code snippets, focusing on security aspects.
They found that developers often select code snippets based on additional information, such as response score and acceptance status, without considering the security of the snippet.
Reusing code snippets from Stack Overflow is common also among experienced developers.
However, on average, software that utilizes such snippets has a higher percentage of bugs after code reuse~\citep{RN13870}.
\cite{DBLP:journals/ese/BaltesD19} found that developers frequently reuse content from Stack Overflow without the required attribution, making it difficult or even impossible to go back to the original post in case problems arise.
Many Stack Overflow code snippets are incomprehensible, of low quality, or require extensive customization for effective reuse~\citep{DBLP:journals/ese/WuWBI19}.
Consequently, the following section examines the factors that contribute to the readability and comprehensibility of source code.

\subsection{Code reading and comprehension}
\label{sec:code-comprehension}

Source code is a set of computer instructions written in text form using a language that can be understood by humans.
Developers often face time-sensitive situations that require rapid product development or bug fixes.
In such cases, they rely on existing code to solve their problems~\citep{RN13871}.
Code transparency refers to its readability and reproducibility.
In a study by \cite{RN13871} on transparency, programmers emphasized the importance of elements such as organization, style, and architecture.
Among these elements, code style, including comments, was particularly emphasized.
Code comments are added to improve readability and support future maintenance activities.
The usefulness of comments depends on the programming experience of the reader~\citep{RN13871}.
Program readability, which refers to the process by which developers understand source code, is a critical aspect of software development.
The readability of code differs from that of human-readable text or natural languages due to its structured nature and the inclusion of various elements such as design, documentation, and logic~\citep{RN13879}.
The higher the readability of the code, the faster and more accurately a programmer can extract critical information from the program text.

The literature on code comprehension has identified different approaches to reading code.
Researchers have proposed several cognitive models~\citep{fagerholm2022cognition} of the reading process to explain how software developers approach the process of code comprehension.
These models include \emph{top-down}, \emph{bottom-up}, \emph{as-needed}, and \emph{integrated comprehension} approaches~\citep{RN3810}.

\cite{RN13878} has proposed the \emph{top-down theory} of code comprehension.
According to this theory, software developers begin by forming general hypotheses about the purpose of the code.
In the next step, developers attempt to verify these hypotheses by scanning the code for specific structures or operations, which \cite{RN13878} calls ``beacons'' that programmers use to recognize known structures and operations in the code.
According to \cite{RN13878}, code and its documentation can be thought of as a collection of beacons, such as comments, variables, or indentation.
When these indicators are found, they confirm the presence of expected structures or operations and validate the developer's initial hypotheses.
However, if no indicators are found, the programmer must examine the code more closely and may need to revise or reject their hypotheses.
Understanding continues in iterative rounds, in which software developers repeatedly formulate and verify hypotheses until they have a complete understanding of the entire code \citep{RN13878, RN3810}. 

The \emph{bottom-up theory} suggests that software developers start by examining small pieces of code, gradually combining them to form higher levels of abstraction, as their initial knowledge is insufficient to identify beacons~\citep{RN13996}.
In this approach, semantic relationships are grouped into chunks that serve as building blocks for higher-level chunks.
The process continues as more chunks are constructed and relationships between existing chunks are recognized, leading to the formation of larger, higher-level chunks. This iterative process helps the software developer form a general hypothesis about the purpose of the program~\citep{RN13996, RN3810}.

\cite{RN13987} proposed the \emph{as-needed approach}, in which software developers aim to understand only the parts of the code necessary to make successful changes, thereby minimizing the amount of code they need to understand. This strategy focuses on selectively understanding specific components rather than the entire code. 

The \emph{integrated model of code comprehension} combines elements of top-down and bottom-up strategies and suggests that software developers alternate between these strategies depending on the specific context or their familiarity with the code~\citep{RN13988}.
When the code is familiar, developers tend to use a top-down approach.
In contrast, in scenarios where the code is unfamiliar, such as when reading code snippets on Stack Overflow, the bottom-up approach is more frequently used~\citep{RN3810}.

Code style is highly relevant for code readability.
Several programming languages provide style guides that outline coding conventions and serve as standard libraries for writing code.
For the Python programming language, \emph{PEP 8 - Style Guide for Python Code} describes coding conventions to ensure code consistency~\citep{RN14007}.
Violations of the style guide have a negative impact on code readability~\citep{RN13986}. Moreover, Stack Overflow posts with Python code snippets that adhere to the coding style guide and have fewer violations per statement are preferred by the community \citep{RN13873}. 

Beyond code style, code comments, which are the main focus of this paper, are particularly relevant for code comprehension.
\cite{RN13993} have highlighted the importance of source code comments as an information source for software developers.
Comments help improve the readability, comprehensibility, and maintainability of the code~\citep{RN13993}.
Source code and comments are the two most important artifacts for understanding a system~\citep{RN13998}.

More recently, software engineering researchers have studied aspects such as the effect of functional decomposition on code comprehensibility, without finding a connection~\citep{DBLP:conf/iwpc/Tempero0FLK0SST24}.
\citet{DBLP:conf/iwpc/SergeyukLTSBKB24} have studied whether existing code readability models are aligned with developers' comprehensibility ratings for AI-generated code, considering the models proposed by \citet{DBLP:conf/issta/BuseW08}, \citet{DBLP:conf/msr/PosnettHD11}, Dorn (2012)\footnote{\url{https://web.eecs.umich.edu/~weimerw/students/dorn-mcs-paper.pdf}}, \citet{DBLP:journals/jss/MiHOM22}, and \citet{DBLP:journals/smr/ScalabrinoLOP18}.
However, they found that readability assessments of AI-generated code differed between these models and that the correlation with human comprehensibility evaluations was low.
\citet{DBLP:conf/iwpc/FakhouryRHA19} found that existing readability models do not capture readability improvements as documented in open source GitHub projects.
\citet{DBLP:conf/iwpc/EtgarFHPF22} have studied the connection between the information contained in function and variable names and their memorability. They found that more informative names are easier to remember.

Moreover, researchers have studied ``atoms of confusion'', that is, small patterns of source code that might be misinterpreted, in languages such as  C/C++~\citep{DBLP:conf/msr/GopsteinZFC18} and Java~\citep{DBLP:conf/iwpc/LanghoutA21}.
\citet{DBLP:conf/iwpc/StapletonGLEWL020} studied the comprehensibility of human-generated and machine-generated code summaries and found that their participants performed significantly better when using human-written summaries.
\citet{DBLP:conf/icse/WieseRF19} found that code written using ``expert'' patterns is considered more elegant, but its comprehensibility is not necessarily different from ``novice'' patterns.

Regarding code comments, \cite{DBLP:conf/scam/JabrayilzadeGT21} developed a taxonomy of 11 inline comment smells based on a multivocal literature review and 899 inline comments from three open-source Java projects that the authors manually labeled.
The most common smells were misleading comments that did ``not accurately represent what the code does'' and obvious comments that ``restate what the code does in an obvious manner''~\citep{DBLP:conf/scam/JabrayilzadeGT21}.
Based on an online survey with 102 participants, \cite{DBLP:journals/tosem/HuangGDSCLZZ23} found that inline comments are written more frequently than block comments.
Regarding perceived helpfulness, the authors were unable to find a considerable difference.
With respect to automatic comment generation, they found that existing models perform better for block comments.
Finally, \cite{DBLP:journals/tosem/WangGBGWWY24} found that TODO comments in open-source projects are often of low quality.

\subsection{Summary}
\label{sec:summary-related-work}

Although there is a considerable body of knowledge on different properties of Stack Overflow posts (see \autoref{sec:stack-overflow}) and code comprehension in general (see \autoref{sec:code-comprehension}), the impact of different ways of presenting and documenting source code in the typically short code snippets on Stack Overflow has not yet been studied.
Stack Overflow's fragmented code-snippet-based knowledge sharing approach differs from traditional code comprehension, which usually spans larger code bases, potentially even large industrial software projects.
For example, results from whole open-source projects regarding the differences between inline and block comments do not necessarily generalize to individual code snippets in Stack Overflow posts.
In the specific context of Stack Overflow, users' assessment of the perceived helpfulness of code snippets can influence their decision to reuse.
A better understanding of perceived helpfulness has implications beyond code on Stack Overflow, as AI-based assistants also offer the option to generate multiple solution proposals or re-generate solutions that users are not satisfied with.

\section{Hypotheses}

In the following, we describe the hypotheses that we formulated based on our research questions. These hypotheses guided the design of our study.

\subsection{Code comments and inline vs. block comments (RQ1)}

The importance of code comments is supported by the study by \cite{RN13990}, who analyzed the relationship between code comments in Python code and problem solving through various experiments.
The study concluded that comments play an important role in problem solving within coding projects and that increasing the percentage of relevant comments, along with source code comments, can reduce the average time it takes to solve a problem.
Previous studies also support the claim that code comments play a crucial role in the selection of code snippets on Stack Overflow.
\cite{RN13992} conducted a study on Stack Overflow, examining well-received answers and identifying characteristics of effective examples.
Their findings highlighted the importance of code explanations in addition to code snippets, with comments recognized as being as important as code snippets~\citep{RN13992}.
Moreover, \cite{RN13995} conducted a survey of professional developers and found that they value well-structured and complete documentation, as well as a comprehensive set of examples.
Users are likely to subjectively perceive greater helpfulness when they encounter code snippets accompanied by comments on Stack Overflow.
Based on the previous findings, we formulated the following hypothesis:

\begin{hypothesis}{1a}
Code snippets with code comments are generally rated as more helpful than code snippets without code comments.
\end{hypothesis}

In the Python programming language, comments are preceded by a hash character (\texttt{\#}) and a space.
For the context of this study, we ignore multi-line comments and Python docstrings.
We distinguish two types of comments used within code snippets: block comments and inline comments.
Block comments refer to a sequence of consecutive comment lines, each of which begins with a hash sign.
These block comments are similar to method comments~\citep{RN13826, DBLP:journals/tosem/WangGBGWWY24}.
Inline comments, on the other hand, appear on the same line as the statement~\citep{RN14007}.
Although there is a large body of literature on code comments in general~\citep{RN13826}, only a few studies have specifically compared different types of code comments in terms of their impact on code comprehension and perceived helpfulness. 
Due to their structural and contextual advantages, block comments are better suited for providing more detailed information and longer explanations than inline comments.
Typically placed at the beginning of a section of code, such as a function, class, or module, block comments can provide a comprehensive overview or summary.
Their advantage is that they are not limited to the space at the end of a line of code, allowing for more elaborate explanations, including multiple paragraphs.
In contrast, inline comments, appropriately positioned next to specific lines of code, are ideal for concise explanations or clarifications and contain fewer words on average~\citep{DBLP:journals/tosem/WangGBGWWY24}.
We suggest that block comments, with their capacity for extended detail, are generally more advantageous than the brevity offered by inline comments.
We can draw on the code comprehension literature to support this claim (see, e.g., \cite{RN3810}).
While providing only inline comments is likely to be effective for detailed, line-specific explanations that support bottom-up and as-needed comprehension strategies, block comments can additionally provide general overviews and support top-down comprehension.
Because code snippets on Stack Overflow are typically unfamiliar to readers, the bottom-up comprehension approach is often required~\citep{RN3810}; the as-needed approach is likely to be applied as well.
Thus, inline comments provide immediate, line-specific explanations that help piece together the lower-level functionality that is essential to the bottom-up approach.
However, for a deeper understanding of the entire proposed code solution, block comments can help explain the overarching design and purpose of code sections, not just the intentions behind specific lines or statements, which inline comments provide.
In addition to code comprehension, users may perceive detailed and carefully crafted block comments as more ``polite'' than brief inline comments.
This perception of politeness is an important factor in answer selection on Stack Overflow~\citep{RN14011}.
In summary, we hypothesize that: 

\begin{hypothesis}{1b}
Code snippets with block comments are generally rated as more helpful than code snippets with inline comments.
\end{hypothesis}

\subsection{Interaction effect with expertise---experts vs. novices (RQ2)}

Novices and experts exhibit differences in their code-reading behavior, so it is important to consider this distinction in expertise when formulating hypotheses about users' preferences for code comments and the specific types of comments they prefer on Stack Overflow.
A study conducted by \cite{RN13974} used eye-tracking to investigate code reading skills and found that novices read code in a less linear manner compared to natural language text.
Additionally, experts were found to read code even less linearly than novices, indicating a difference in the reading process between code and natural language text.
In examining how subjects view an algorithm, \cite{RN13977} found that the eye movements of experts and novices differed when viewing the English and Pascal versions of an algorithm.
They found that experts spent more time looking at complex statements.
They also discovered that while both groups spent a lot of time looking at comments, novices spent significantly more time looking at comments~\citep{RN13977}, which implies that comments seem to be more important to novices.
Learning a programming language involves the acquisition of knowledge structures to apply problem-solving skills~\citep{RN13983}.
Programming expertise involves different cognitive processes that, combined with changes in knowledge, can lead to the adoption of different methods for solving specific programming problems.
These different approaches can be described as strategies, and experienced programmers have a more extensive repertoire of strategies than novices when programming~\citep{RN13983}.
An exploratory study conducted by~\citep{RN13975} aimed to understand how novice software engineers focus on the information presented in responses to questions on Stack Overflow.
The results showed that novice software engineers only pay attention to 27\% of the code and 15 to 21\% of the text in a Stack Overflow post as they try to understand the relevant information and determine how to apply it to their specific context.
As \cite{RN13993} discuss, code comments serve as a valuable source of information.
\cite{RN13983} argues that the choice of a code snippet is influenced by the developer's knowledge.
Based on the existing literature, which suggests that novice software developers benefit from more explanatory comments, while experienced programmers tend to prefer fewer or no source code comments, we propose the following hypothesis:

\begin{hypothesis}{2a}
Individuals with less programming experience perceive code snippets with comments as more helpful than those without comments.
\end{hypothesis}

Furthermore, we argue that block source code comments have the potential to provide novice software developers with the additional information they need compared to shorter inline comments.
In contrast, experienced programmers tend to prefer inline source comments, which are typically shorter and cause less disruption to the code-reading process. Based on these considerations, we formulate the following hypothesis:

\begin{hypothesis}{2b}
Individuals with less programming experience perceive code snippets with block comments as more helpful than those with inline comments.
\end{hypothesis}

\subsection{Ordering effects (RQ3)}

The order of answers within a Stack Overflow thread can influence their perceived helpfulness.
Typically, Stack Overflow sorts answers in descending order by answer score.
\cite{DBLP:journals/jcmc/MurphyHM06} observed that an item's position affects its memorability, with top items enjoying a memory advantage due to early processing and less competition from subsequent items, a phenomenon known as the primacy effect.
In support of this, \cite{RN13981} argued that individuals tend to process horizontal lists from top to bottom, resulting in longer memory retention for initial options.
Their studies showed that participants looked longer at the first option on a list, especially for longer lists.
This suggests that due to cognitive and behavioral biases, code options at the beginning of a list are more likely to be selected, regardless of code quality or comment type. This leads to the hypothesis that:

\begin{hypothesis}{3}
Regardless of the content, answers positioned earlier in the thread are generally perceived as more helpful by users.
\end{hypothesis}

\subsection{Answer score (RQ4)}

The Stack Overflow platform uses a scoring system for both questions and answers, allowing users to vote on the helpfulness of a question or answer.
Several studies have looked at different aspects of the Stack Overflow community and platform, such as the accepted answer score, the scoring system, and the user reputation.
For example, \cite{RN13982} studied Stack Overflow posts and explored the interaction between accepted answers and upvotes.
Their results showed that accepting an answer led the community to vote for that answer instead of an alternative answer with a similar or higher score prior to acceptance.
\cite{RN14003} conducted the first large-scale between-groups experiment and observed that answer details, called surface features, have a significant impact on the selection of both secure and insecure code snippets.
The observed positive effect of the answer score on the acceptance and selection of code snippets also has its drawbacks.
Due to Stack Overflow's decentralized structure, its voting and reputation system may be susceptible to manipulation~\citep{RN14009}.
Answer details, such as the `accepted answer' mark, user reputation scores, and answer scores, have also been found to mislead users into accepting insecure coding advice, inadvertently promoting vulnerable code~\citep{RN13989}.
Building on previous studies that highlight the importance of answer scores in code snippet selection, the following hypothesis was formulated regarding the perceived helpfulness of code snippets:

\begin{hypothesis}{4}
Users perceive answers as more helpful when they not only appear earlier in the thread, but also have a higher answer score, regardless of their content.
\end{hypothesis}

\section{Method}

Our empirical study utilized an online experiment that simulated the Stack Overflow environment, followed by a post-survey.
Participants were asked to rate answers based on their perceived helpfulness within a simulated Stack Overflow environment.
The \emph{answer position} and \emph{source code comment type} (block comment, inline comment, no comment) were used as within-subjects factors in the experimental design.
In addition to \emph{answer position}, the experiment used the \emph{answer score} as a between-subjects factor, with one group receiving the score treatment, displayed on the left side of the Stack Overflow answers, while the control group received the answers in the same order, but without scores.
Participants were randomly assigned to either the treatment or control group. This experimental setup allowed us to isolate the effect of scores from content-based factors such as code comments, enabling a controlled comparison of perceived helpfulness across conditions.
For each scenario, that is, for each Stack Overflow question, the authors developed three different code snippets, inspired by real Stack Overflow posts related to the question.
Each answer consisted of a short explanation text and a source code snippet.
We counterbalanced the comment types with source code snippets.
Since each participant was presented with three code snippets for each question, this resulted in six possible combinations that were randomly selected to be presented to a participant for each scenario.
To vary the order of the answers, we presented the three answers to the participants in random order.

\subsection{Experimental materials: Stack Overflow simulation}

In the experiment, participants were instructed to imagine searching for answers on Stack Overflow: \emph{Suppose that you are developing a Python program and run into a specific programming problem. While searching for your problem on the web, you find an answered thread on Stack Overflow that addresses your problem.}
Participants were instructed to read through all available answers, along with accompanying code snippets, and then rate each answer based on its helpfulness.

\begin{figure}
    \centering
    \includegraphics[width=0.7\textwidth]{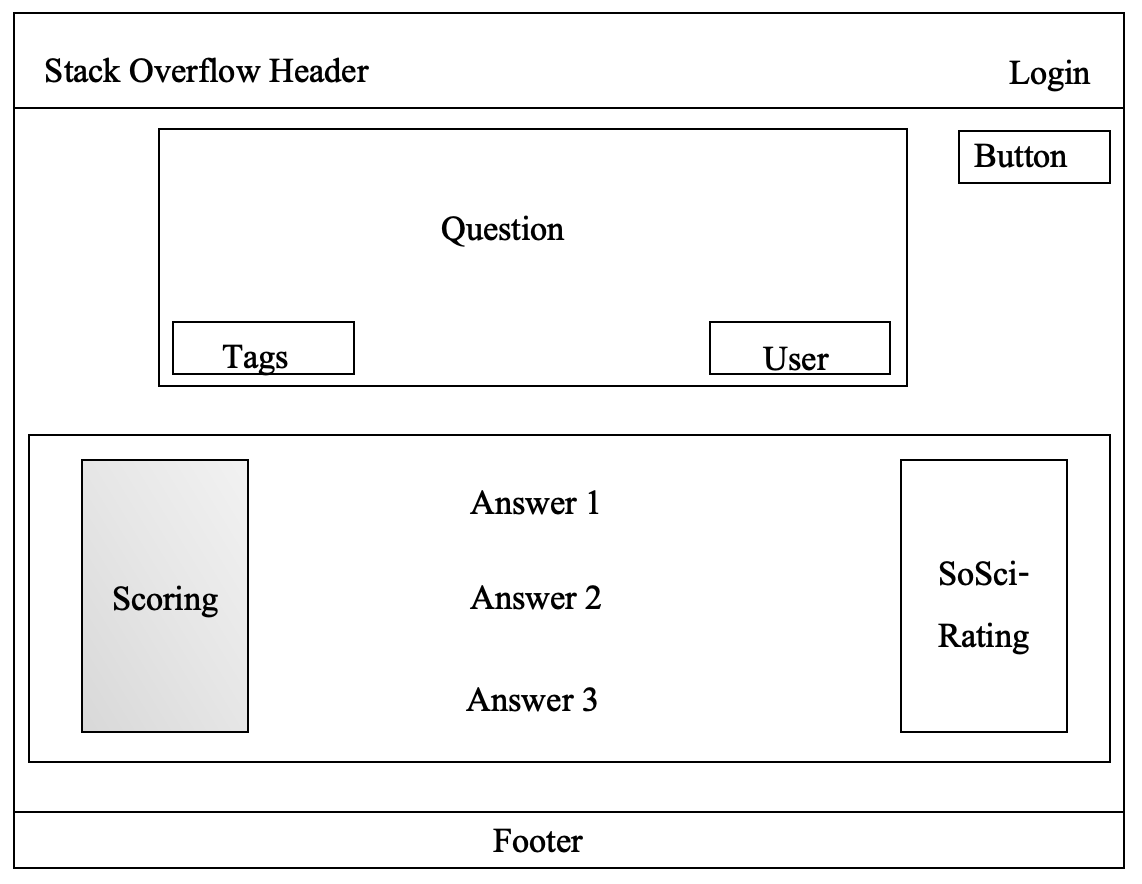}
    \caption{Simulated Stack Overflow design (abstract structure).}
    \label{fig:so-design-abstract}
\end{figure}

\begin{figure}
    \centering
    \fbox{\includegraphics[width=0.98\textwidth]{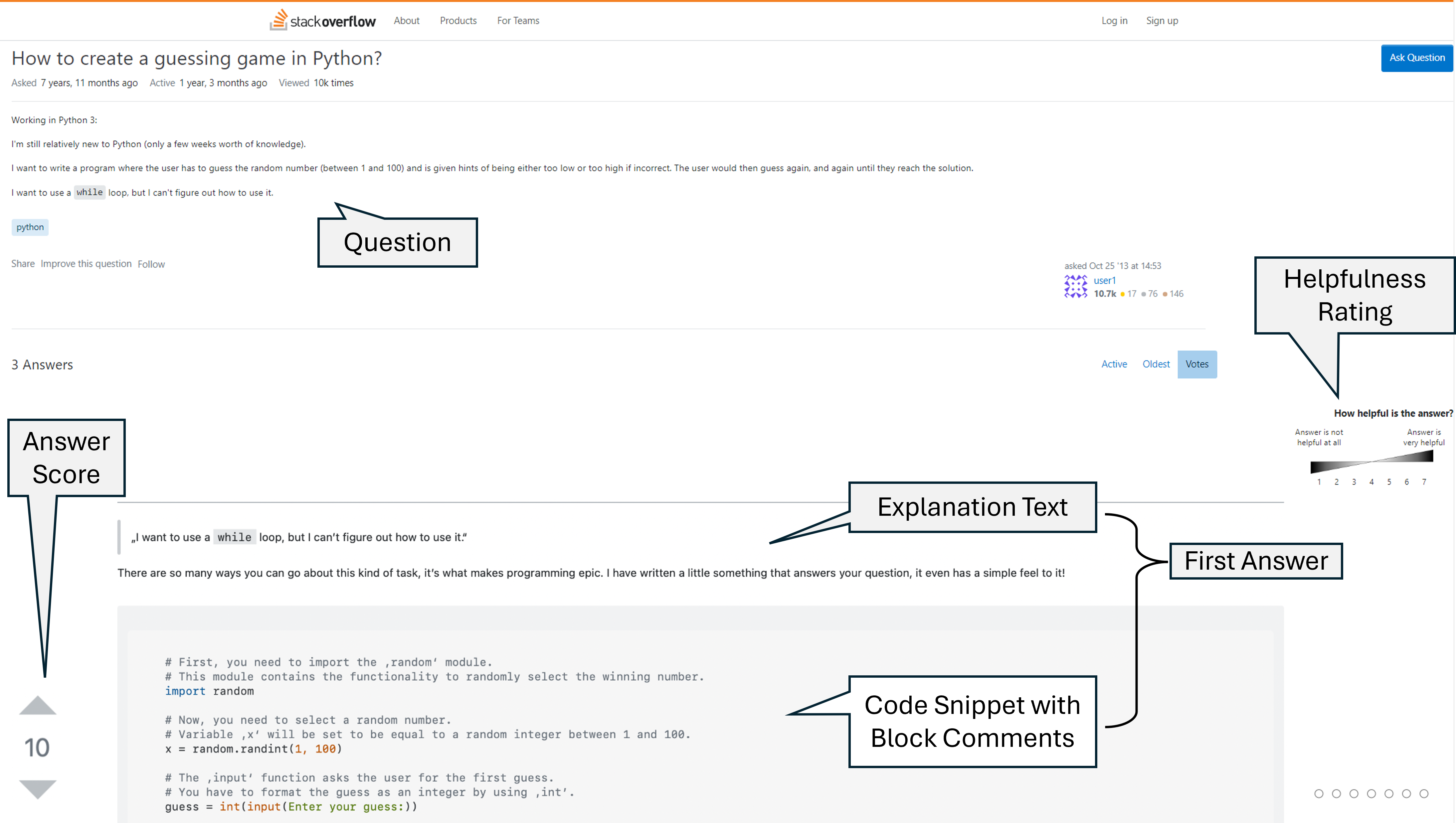}}
    \caption{Simulated Stack Overflow design (labeled screenshot).}
    \label{fig:so-design-screenshot}
\end{figure}

\begin{figure}
    \centering
    \includegraphics[height=2cm]{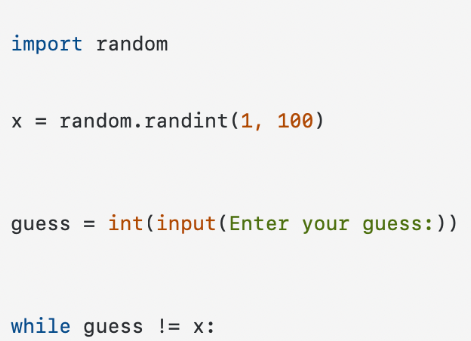}
    \hfill
    \includegraphics[height=2cm]{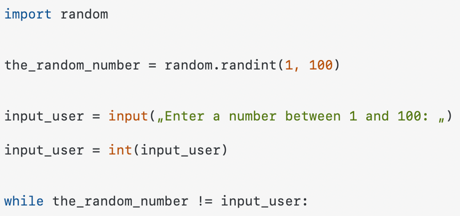}
    \hfill
    \includegraphics[height=2cm]{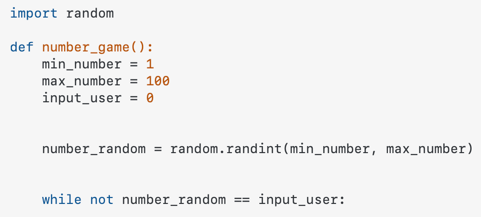}
    \caption{Excerpts from three answers without source code comments.}
    \label{fig:excerpts-no-comment}
\end{figure}

\begin{figure}
    \centering
    \includegraphics[width=\textwidth]{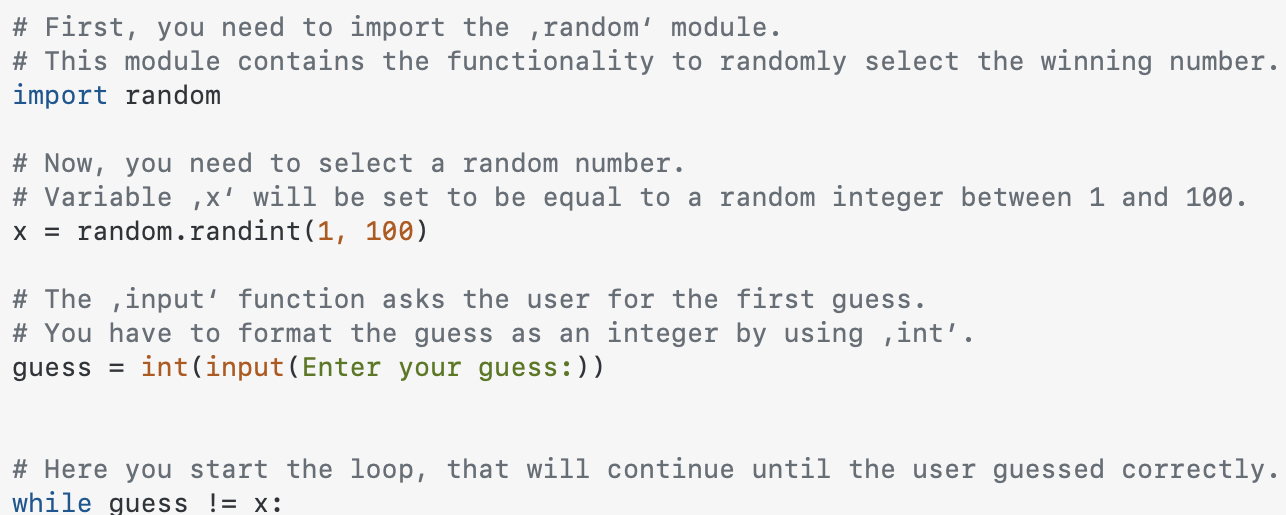}
    \caption{An excerpt from an answer with block source code comments.}
    \label{fig:excerpt-block-comment}
\end{figure}

\begin{figure}
    \centering
    \includegraphics[width=\textwidth]{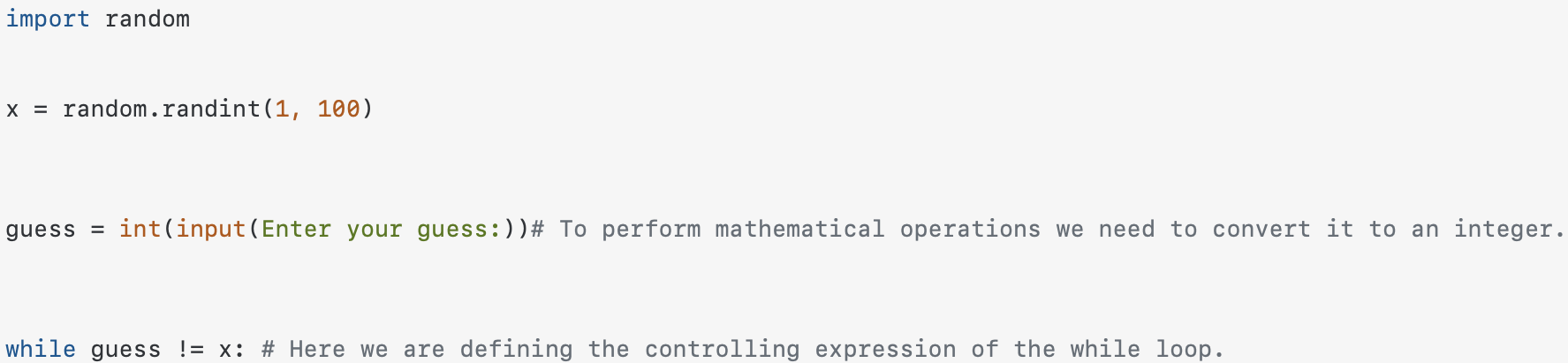}
    \caption{An excerpt from an answer with inline source code comments.}
    \label{fig:excerpt-inline-comment}
\end{figure}

The \emph{Stack Overflow simulation} consisted of three Stack Overflow thread scenarios, each with the same structure and layout (as depicted in Figure~\ref{fig:so-design-abstract}), inspired by existing posts on the platform.
Rather than randomly sampling threads, we constructed three Python-related scenarios based on real Stack Overflow discussions, but adapted them to ensure a uniform style and a manageable level of complexity. Each set of code snippets was designed to compile and produce similar outcomes, employing various functions, variables, or packages. 
Each scenario began with a question posed to the forum by a fictitious Stack Overflow user, representing a specific problem.
Figure~\ref{fig:so-design-screenshot} shows the first Stack Overflow scenario with the question \emph{``How to create a guessing game in Python?''} and the corresponding problem of the Stack Overflow user identified as \emph{``user1''}, along with the first answer.
The experiment included three different Stack Overflow scenarios, one asking how to print the fourth most common word from a \emph{``Romeo and Juliet''} excerpt using Python, and another asking for guidance on how to handle exceptions in Python.
We developed questions and answers for the threads using Python because it is among the most popular programming languages on Stack Overflow~\citep{RN14008}.
The code snippets were developed in a way that each code snippet yields the same or very similar results.
The question was followed by three answers that provided a solution to the problem posed.
Figure~\ref{fig:excerpts-no-comment} shows snippets from three answers from the control group without source code comments for the first Stack Overflow scenario to the question \emph{``How to create a guessing game in Python?''}.
Figure~\ref{fig:excerpt-block-comment} provides an excerpt from one of the answers with block source code comments, and Figure~\ref{fig:excerpt-inline-comment} shows an excerpt from one of the answers with inline source comments.

For the treatment group with answer scores, scoring buttons were displayed on the left side (see Figure~\ref{fig:so-design-screenshot}).
The answers were displayed to the participant in descending order of the scores:
starting with $10$, the highest score, followed by a score of $3$, and finally a score of $0$.
These scoring numbers are realistic because the average score for an answer is $1.8$~\citep{RN13992}.

After reading the scenario and answer options, participants were asked to rate the helpfulness of each answer option.
This was done using a 7-point Likert scale item ranging from ``Answer is not helpful at all'' to ``Answer is very helpful.''
The scale was prominently displayed to the right of each answer code snippet (see Figure~\ref{fig:so-design-screenshot}).

\section{Experimental procedure and post-survey}

The online experiment was implemented using the questionnaire platform SoSci\footnote{\url{https://www.soscisurvey.de/}} and was divided into three main parts: the introduction, the Stack Overflow simulation, and the post-survey.
In the first part, the introduction, the participant was presented with a consent form that explicitly stated that participation was voluntary and anonymous, and that all data collected would be kept confidential.
In addition, participants were given technical instructions, such as the requirement to use a PC or laptop (not a mobile device) and to ensure that the browser zoom was set to 100\%.
After the Stack Overflow simulation, the third part was a post-survey divided into three sections: (a) Stack Overflow questions, (b) general control questions related to expertise, and (c) demographic information.
In the first part of the post-survey, participants were asked to rank Stack Overflow elements displayed as cards in order of importance when evaluating an answer.
They used drag-and-drop to position each item on a scale from least to most important. The question was asked twice: First, we asked about the elements used in the experiment, that is, the answer score, blank code (i.e., code without comments), explanation text, inline source comments, and block source comments. By ``explanation text,'' we mean a description of the code placed at the beginning of an answer that is not formatted as an inline or block comment.
Second, we asked about the elements of Stack Overflow in general, which included the same elements as before, along with two additional ones: the accepted answer marked as the best answer by the questioner and the answerer's reputation score.
We asked several questions about the user's prior experience and expertise. Participants were asked about their active or passive experience with Stack Overflow, frequency of use, whether they had used a code snippet from the Q\&A page, and their reasons for using Stack Overflow. They also provided information about their programming skills, including years of experience. Experience with different programming languages was rated on a 6-point Likert scale item from ``never'', ``rarely'', ``seldom'', ``sometimes'', ``occasionally'' to ``most of the time.''
An open text field was provided to list additional programming languages not included. In the last section of the questionnaire, participants provided demographic information such as gender, age, country of origin, education, and occupation or field of study. 

\subsection{Data-collection, sample, and data preparation}

In order to collect empirical data from software development students and professionals, we chose several channels to recruit participants for this online study.
First, the mail delivery service of a large European university was used.
Specific faculties and study programs---such as Mathematics and Computer Science, Physics, and Information Systems---were selected as recipients of the invitation email. An email was sent to subscribers of the survey mailing list who belonged to the selected student groups, including a short explanation of the study and a link to the Stack Overflow online experiment.

The study was also posted on a local hackerspace website and on LinkedIn. In addition, individuals with programming backgrounds from the authors' personal networks---including friends, work colleagues, and family members---were invited to participate in the study to ensure a diverse mix of novice and professional software developers.

Only fully completed questionnaires were considered for further analysis. The sample consisted of 91 participants with a mean age of 26.22 years ($SD = 7.58$), of whom 79.1\% identified as men, 18.7\% as women, and 2.2\% preferred not to disclose their gender.

To prepare for the analyses and test the hypotheses, we first divided the participants into novices and experts based on their programming skills, as determined by their responses to the post-survey. Novices were those with none, less than one year, or one year of programming experience, while experts had two to ten or more years of experience. Of the 91 participants, $61$ (67\%) were classified as experts and $30$ (33\%) as novices.

This imbalance can be attributed to the data collection process, as most respondents were recruited through the university mailing list---where many computer science students already have considerable programming experience---and through LinkedIn or colleagues, where we aimed to recruit professional software developers or those with similar backgrounds.

\section{Results}

In the following, we summarize our results on the helpfulness of different types of source code comments in Stack Overflow posts, as well as the perceived importance of various Stack Overflow elements.

\subsection{Helpfulness of source code comment types (RQ1 and RQ2)}

\begin{figure}
    \centering
    \includegraphics[width=0.8\textwidth]{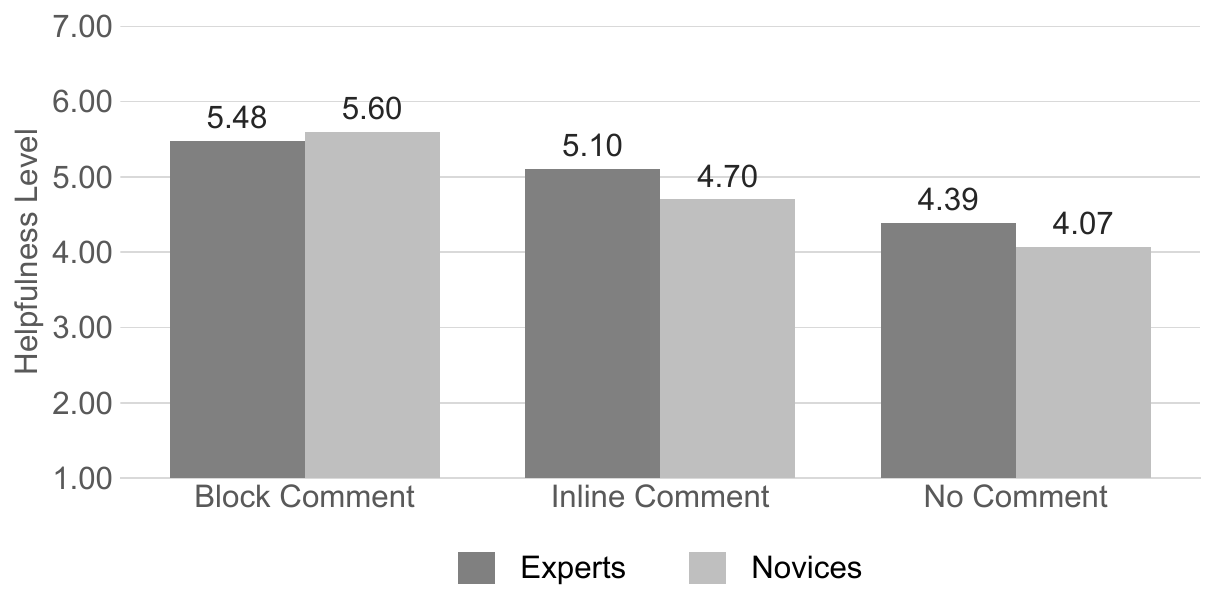}
    \caption{Comparison of the mean helpfulness of different types of source code comments.}
    \label{fig:comparison-average-helpfulness2}
\end{figure}

The analyses are based on the cumulative sample of three answer comments across three scenarios per participant, resulting in a total of $n_\text{accumulated} = 819$ observations of answer comment helpfulness ratings ($n = 91 \times 3 \times 3$). 

Figure~\ref{fig:comparison-average-helpfulness2} shows how programming novices and experts rated the helpfulness of different types of source code comments. Novices rated block comments as slightly more helpful ($M = 5.60$, $SD = 1.66$) than experts ($M = 5.48$, $SD = 1.50$), while the average ratings for inline comments and no comments were lower for novices than for experts. Specifically, experts rated inline comments ($M = 5.10$, $SD = 1.42$) as more helpful than novices ($M = 4.70$, $SD = 1.62$). Experts also rated uncommented code as more helpful on average ($M = 4.39$, $SD = 1.63$) than novices ($M = 4.07$, $SD = 1.62$). Overall, both groups rated commented code as more helpful than uncommented code snippets.

We used linear mixed models provided by the \texttt{lme4}\footnote{\url{https://cran.r-project.org/web/packages/lme4/}} and \texttt{lmerTest}\footnote{\url{https://cran.r-project.org/web/packages/lmerTest/}} packages in R\footnote{\url{https://www.r-project.org}} to analyze the fixed factors: comment type (baseline = no comments, block comments, inline comments), expertise (experts vs.\ novices), position (first, second, third answer), and the presence of answer scores (no answer score vs. presence of answer score; first answer: score $10$, second answer: score $3$, and third answer: score $0$). The dependent variable was the perceived helpfulness of an answer. We also examined interaction effects between comment type and expertise, as well as between position and score.

A second model (`Model 2' in Table~\ref{tab:table1}) was calculated to allow for a more direct comparison of block and inline comment types. This model excluded the ``no comment'' condition and used block comments as the baseline. Both models included two random intercepts, one for participant and one for code snippet. The intra-class correlation (ICC) was $0.23$ in the first model and $0.24$ in the second model, indicating that a substantial proportion of the variance in helpfulness ratings is explained by differences between participants and code snippets. The conditional $R^2$, reflecting the variance explained by both fixed and random effects, was $0.302$ for the first model and $0.269$ for the second, indicating that approximately 30\% and 27\% of the variance, respectively, could be explained by the model.

The position of the answer, the presence of an answer score, and their interaction did not have a significant effect on the perceived helpfulness of a code snippet in either model.
Expertise was also not a significant factor; thus, novices did not generally rate the helpfulness of comments higher than experts.

Furthermore, the position of the comment (first, second, or third answer) had no significant influence on helpfulness ratings.

The results of Model 1 suggest that including block comments (Estimate = 1.51, $p < 0.001$) or inline comments (Estimate = 0.62, $p = 0.003$) significantly increases the perceived helpfulness of answers compared to uncommented source code.

Notably, novices perceive block comments as particularly helpful in comparison to no comments---even more so than experts---as indicated by a marginally significant interaction effect (Estimate = \ensuremath{-0.43}, \ensuremath{p < 0.09}), as shown in Figure~\ref{fig:comparison-average-helpfulness2}.
This suggests that the effect of block comments on perceived helpfulness depends on the expertise level of the user.

In contrast, the interaction between expertise and inline comments vs. no comments was not significant (Estimate = 0.09, $p = 0.723$).
Model 2 further shows that, when comparing block comments to inline comments, block comments are rated as significantly more helpful (Estimate = \ensuremath{-0.88}, $p < 0.001$).
Furthermore, the interaction between expertise and inline comments vs. block comments shows a significant positive effect on perceived helpfulness (estimate = 0.50, $p = 0.042$). Experts and novices rated block comments as more helpful than inline comments, with the difference being more pronounced for novices (see Figure~\ref{fig:comparison-average-helpfulness2}).

Table~\ref{tab:table1} summarizes the results of the linear mixed-effects models for perceived helpfulness.

\begin{table}
    \caption{Linear mixed models for perceived helpfulness: CT~=~comment type, Position~=~answer position in the thread, User Ratings~=~presence of answer scores, Exp.~=~expertise (experts vs.\ novices), \textbf{*}~denotes interaction effects, bold values mark effects with $p<0.10$.}
    \label{tab:table1}
    \centering
    \footnotesize
    \setlength{\tabcolsep}{3pt}
    \renewcommand{\arraystretch}{1.2}
    \begin{tabular}{p{4.2cm} *{4}{>{\raggedleft\arraybackslash}p{1.5cm}}}
    \toprule
     & \multicolumn{2}{c}{\textbf{Model 1}} & \multicolumn{2}{c}{\textbf{Model 2}} \\
     & \multicolumn{2}{c}{comparing block} & \multicolumn{2}{c}{comparing block comments} \\
     & \multicolumn{2}{c}{and inline comments} & \multicolumn{2}{c}{with inline comments} \\
     & \multicolumn{2}{c}{with no comments} & \multicolumn{2}{c}{} \\
    \midrule
    \textbf{Predictors} & \multicolumn{1}{c}{\textit{Estimates}} & \multicolumn{1}{c}{\textit{p}} & \multicolumn{1}{c}{\textit{Estimates}} & \multicolumn{1}{c}{\textit{p}} \\
    \midrule
    (Intercept) & 4.08 & \textless0.001 & 5.57 & \textless0.001 \\
    CT (Block vs.\ No) & 1.51 & \textbf{\textless0.001} & \multicolumn{2}{c}{\spanval{\textit{not included}}} \\
    CT (Inline vs.\ No) & 0.62 & \textbf{0.003} & \multicolumn{2}{c}{\spanval{\textit{not included}}} \\
    CT (Inline vs.\ Block) & \multicolumn{2}{c}{\spanval{\textit{not included}}} & $-0.88$ & \textbf{\textless0.001} \\
    Expertise & 0.32 & 0.172 & $-0.1$ & 0.674 \\
    Position & 0.02 & 0.777 & 0.05 & 0.524 \\
    User Ratings & 0 & 0.997 & 0.11 & 0.551 \\
    \midrule
    CT (Block vs.\ No) \textbf{*} Exp.\ & $-0.43$ & \textbf{0.09} & \multicolumn{2}{c}{\spanval{\textit{not included}}} \\
    CT (Inline vs.\ No) \textbf{*} Exp.\ & 0.09 & 0.723 & \multicolumn{2}{c}{\spanval{\textit{not included}}} \\
    CT (Inline vs.\ Block) \textbf{*} Exp.\ & \multicolumn{2}{c}{\spanval{\textit{not included}}} & 0.5 & \textbf{0.042} \\
    Position \textbf{*} User Ratings & $-0.03$ & 0.793 & $-0.02$ & 0.897 \\
    \midrule
    \textbf{Random Effects} & & & & \\
    \midrule
    $\sigma^2$ & 1.9 & & 1.77 & \\
    $\tau_{00}$ Participant & 0.46 & & 0.49 & \\
    $\tau_{00}$ Code Snippet & 0.1 & & 0.06 & \\
    ICC & 0.23 & & 0.24 & \\
    \midrule
    N & \multicolumn{4}{c}{\spanval{91 participants}} \\
     & \multicolumn{4}{c}{\spanval{9 code snippets}} \\
    Observations & \multicolumn{2}{c}{\spanval{811}} & \multicolumn{2}{c}{\spanval{539}} \\
    Marginal $R^2$/Conditional $R^2$ & \multicolumn{2}{c}{\spanval{0.099/0.302}} & \multicolumn{2}{c}{\spanval{0.040/0.269}} \\
    \bottomrule
    \end{tabular}
\end{table}

\subsection{Perceived importance of Stack Overflow elements (RQ3 and RQ4)}

Next, we present the results of the post-survey, which assessed participants' perceptions of the importance of various information elements used in Stack Overflow threads. Importance ratings were collected using a fine-grained scale ranging from $0$ to $100$.
Participants were first asked to rate how important each element used in the experiment was when choosing an answer.
A subset of the sample, the participants who had prior experience with Stack Overflow either as active contributors or passive users (85 participants, 93.4\%), were asked to rate these elements again, this time in the context of the real Stack Overflow platform.
In this second assessment, participants were presented with two additional elements that are present on the actual platform but were not included in the experiment: the answerer's reputation score and the accepted answer (i.e., the answer marked as the best by the questioner).

\begin{figure}
    \centering
    \includegraphics[width=\textwidth]{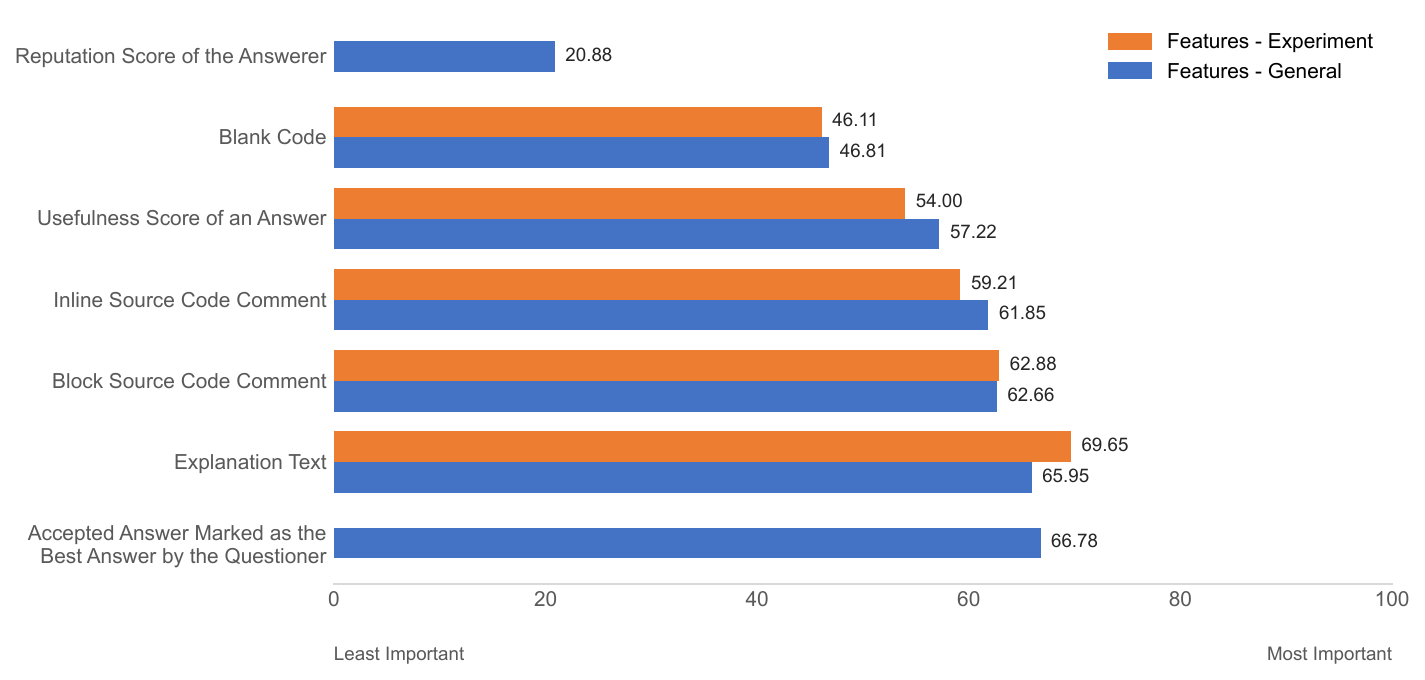}
    \caption{Comparison of average importance ratings of Stack Overflow information elements.}
    \label{fig:comparison-average-importance}
\end{figure}

As shown in Figure~\ref{fig:comparison-average-importance}, the five elements common to both the simulated experimental platform and the real Stack Overflow platform were ranked similarly in terms of perceived importance. Based on descriptive mean values, \textit{explanation text} was rated as the most important element, followed by \textit{block source code comments}. \textit{Inline code comments} ranked third, and the \textit{answer score} ranked fourth. The element rated lowest was \textit{blank code}.
Figure~\ref{fig:comparison-average-importance} also includes ratings for two informational elements that are exclusive to the real Stack Overflow platform and were not part of the experimental interface. Among these, the \textit{accepted answer} was rated as the most important feature for evaluating an answer ($M = 66.78$, $SD = 27.06$), whereas the \textit{reputation score of the answerer} was rated as the least important ($M = 20.87$, $SD = 24.46$).

To statistically assess these differences, a repeated-measures General Linear Model (GLM) was conducted in SPSS on the seven informational elements as perceived in the context of the real Stack Overflow platform. Multivariate tests revealed a statistically significant main effect of the within-subject factor (Pillai's Trace = .779, $F$(6, 79) = 46.54, $p < .001$), indicating substantial differences across the features. 

As Mauchly’s test indicated a violation of sphericity ($W = .208$, $p < .001$), Greenhouse–Geisser correction was applied. The within-subject effect remained statistically significant after correction, $F$(3.93, 330.10) = 31.29, $p < .001$.

Post-hoc Bonferroni-adjusted comparisons showed that the \textit{reputation score} was rated significantly lower than all other features ($p < .001$). Similarly, \textit{blank code} was rated significantly lower than \textit{accepted answers} ($p < .001$), \textit{explanation text} ($p < .001$), and both \textit{inline} ($p = .001$) and \textit{block code comments} ($p = .005$). In contrast, \textit{answer scores}, \textit{accepted answers}, \textit{explanation texts}, and \textit{code comments} (inline and block) did not differ significantly from one another, forming a cluster of consistently high-valued elements. No significant difference was found between inline and block code comments.

A separate repeated-measures GLM was conducted to assess the perceived importance across the five elements used in the experimental platform. This model also revealed a significant main effect of informational element type (Pillai's Trace = .252, $F$(4, 87) = 7.32, $p < .001$). Estimated marginal means indicated that \textit{explanation text} received the highest ratings ($M = 69.6$), significantly outperforming both \textit{answer scores} and \textit{blank code}. Inline and block code comments were again rated positively and did not differ significantly from each other.

Comparing both analyses, the model based on the full feature set (real Stack Overflow platform) yielded stronger statistical effects (Pillai's Trace = .779 vs.\ .252) and a greater number of significant contrasts. This model highlighted particularly negative perceptions of \textit{low reputation} and \textit{blank code}, while further reinforcing the positive evaluations of \textit{explanation text} and \textit{code comments}. Across both analyses, explanations and code comments consistently emerged as the most valued elements in evaluating the quality of answers on Stack Overflow.

\subsection{Interpretation} \label{sec:interpretation}

Table~\ref{table:hypotheses} shows which hypotheses were supported and how they relate to the research questions. 

\textbf{Research Question 1} asked whether code comments influence the perceived helpfulness of code snippets on Stack Overflow. The results of the study supported \textbf{Hypothesis 1a}, which proposed that code snippets with source code comments were generally rated as more helpful than those without comments. In our experiment, code snippets with block and inline comments received higher helpfulness ratings than those without comments.
This finding is consistent with previous research highlighting the importance of source code comments for software developers~\citep{RN13990, RN13993}.

Regarding different comment types, our study showed that block comments were generally rated as more helpful than inline comments, supporting \textbf{Hypothesis 1b}. This supports our argument that block comments---by providing a broader overview essential for understanding the overall design and purpose of a code snippet, especially when encountering unfamiliar code on platforms like Stack Overflow---are more beneficial than shorter inline comments. They also facilitate top-down comprehension of code~\citep{RN3810}.

The ranking of information elements deemed important for choosing an answer was in part consistent with the direct helpfulness ratings observed during the experiment. The mean rating for commented code was higher than that for blank code without comments. However, the results did not support our expectation that block comments would be rated as more important than shorter inline comments when evaluating an answer, as both were evaluated similarly.

\textbf{Research Question 2} explored whether expertise (novices vs. experts) moderates the impact of comments on perceived helpfulness. This was tested through \textbf{Hypotheses 2a and 2b}. 
Although novices did not generally rate source comments as more helpful than experts—leading to a rejection of \textbf{Hypothesis 2a}, they did rate block comments as more helpful than inline comments, supporting \textbf{Hypothesis 2b}. This finding confirms that novices may indeed require more detailed information from source comments than experts.

\textbf{Research Question 3} examined whether the position of an answer in a Stack Overflow thread influences its perceived helpfulness when no additional cues such as answer scores are shown. 
The related \textbf{Hypothesis 3}, which proposed that earlier answer positions would increase the perceived helpfulness of answers, was rejected. The rejection of \textbf{Hypothesis 3} contradicts research on the primacy effect, which has shown that options placed at the top are more likely to be selected, regardless of their content~\citep{RN13981}.
One possible interpretation is that both primacy and recency effects~\citep{DBLP:journals/jcmc/MurphyHM06}, which are commonly observed in psychology, may have influenced user ratings.
Another plausible explanation for the lack of effects of position and score could be the small number of response options (only three), which allowed participants to thoroughly read the texts, code, and code comments and incorporate them directly into their helpfulness ratings without relying on heuristics.

\textbf{Research Question 4} asked whether the presence of answer scores would amplify a potential position effect. The related \textbf{Hypothesis 4}, which predicted that highly rated answers in higher positions would be perceived as more helpful, was also rejected. Furthermore, the answer score was rated as relatively less important in the post-survey ranking of information elements, which was consistent with its lack of influence on the observed helpfulness ratings in the experimental data. These results contradict those of \cite{RN14003}, who demonstrated in an empirical study that surface features such as upvotes and other factors influence the selection of a code snippet.

The rejection of \textbf{Hypotheses 3 and 4} is insightful in that it indicates that users prioritize the quality of content over answer scores and position on Stack Overflow. Participants in our study based their judgments primarily on the content of the code and accompanying explanations, rather than on surface-level cues such as answer order or answer scores. This critical approach helps mitigate risks such as manipulated scores~\citep{RN14009} or the acceptance of unsafe coding advice associated with misleadingly high answer scores~\citep{RN13989}.

\begin{table}[ht]
\caption{Summary of research questions, hypotheses, and acceptance}
\label{table:hypotheses}
\centering
\footnotesize
\setlength{\tabcolsep}{3pt}
\renewcommand{\arraystretch}{1.2}
\begin{tabular}{p{3.1cm} l p{5.4cm} l}
\toprule
\textbf{Research Question} & \textbf{H} & \textbf{Description} & \textbf{Acceptance} \\
\midrule
\multirow{2}{=}{\textbf{RQ1:} What is the impact of code comments on the perceived helpfulness of code snippets on Stack Overflow?}
 & \textbf{H1a} & Code snippets with code comments are generally rated as more helpful than code snippets without code comments. & accepted \\
\cmidrule(l){2-4}
 & \textbf{H1b} & Code snippets with block comments are generally rated as more helpful than code snippets with inline comments. & accepted \\
\midrule
\multirow{2}{=}{\textbf{RQ2:} Does the impact of code comments on perceived helpfulness differ between novices and experts?}
 & \textbf{H2a} & Individuals with less programming experience perceive code snippets with comments as more helpful than those without comments. & rejected \\
\cmidrule(l){2-4}
 & \textbf{H2b} & Individuals with less programming experience perceive code snippets with block comments as more helpful than those with inline comments. & accepted \\
\midrule
\textbf{RQ3:} Does the position of a Stack Overflow answer within a Stack Overflow thread affect perceived helpfulness? & \textbf{H3} & Regardless of the content, answers positioned earlier in the thread are generally perceived as more helpful by users. & rejected \\
\midrule
\textbf{RQ4:} Does the presence of answer scores amplify the effect of position on perceived helpfulness? & \textbf{H4} & Users perceive answers as more helpful when they not only appear earlier in the thread, but also have a higher answer score, regardless of their content. & rejected \\
\bottomrule
\end{tabular}
\end{table}

\section{Limitations and threats to validity}

Although this study provides valuable insights into how code comments affect the perceived helpfulness of Stack Overflow answers, several limitations need to be acknowledged.
We organize these limitations along the standard validity categories of external, internal, construct, and conclusion validity. 

\subsection{External validity}

First, the experiment was conducted in a simulated Stack Overflow environment. Although care was taken to closely mirror the platform's design and functionality, participant behavior might differ from real-world use where additional contextual factors apply.
Our design allowed us to isolate the effect of comment types and surface features such as position and answer score. However, it may have introduced artificiality, for example, by fixing answer scores (10/3/0), omitting answer scores in the control group, or not including other influential features such as the accepted answer badge or common sorting mechanisms (e.g., newest first). As a result, the study does not fully capture the broader dimensions of perceived helpfulness of code comments and user behavior present on the real Stack Overflow platform.

Second, the study relied on a convenience sample recruited through university mailing lists, a local hackerspace, LinkedIn, and the authors' personal networks.
Consequently, the exact response rate is unknown, and the sample may not be representative of the broader population of Stack Overflow users.
Although we aimed for heterogeneity by including both students and professionals, the majority of participants were male, which may introduce a bias in the generalizability of the findings.

Third, all code snippets were constructed to emphasize a dominant comment type (block, inline, or none), and we avoided mixing comment types in the same snippet.
Although this allowed us to examine the effect of distinct styles, it may not reflect real-world practices where mixed comment styles are common.

Moreover, the experiment focused exclusively on Python code snippets and a limited number of programming tasks. Findings may therefore not generalize across different programming languages, problem types, or technical domains.

\subsection{Internal validity}

Although the experiment tested various comment types and surface cues in a controlled setting, the artificial nature of the environment may have influenced participant judgments. For example, answer scores were fixed or omitted, and other influential elements such as accepted answer badges or user reputation were not shown. These simplifications, while necessary for experimental control, may have altered the natural evaluation behavior of the participants, introducing potential confounding factors.

\subsection{Construct validity}

Helpfulness ratings reflect subjective perceptions rather than actual reuse behavior or code comprehension. 
Although perceived helpfulness is a meaningful metric, it may not fully capture actual code comprehension or reuse behavior.
Future studies could incorporate complementary methods such as eye-tracking, think-aloud protocols, or comprehension tests to gain more direct insights into how code comments influence understanding and decision-making.

\subsection{Conclusion validity}

The use of a convenience sample and the absence of a known response rate limit our ability to draw generalizable statistical inferences. Although we employed linear mixed-effects models to account for individual- and item-level variance, the moderate sample size and the imbalance between experts and novices may have limited the detection of smaller interaction effects. Future work should aim to replicate these results in larger and more diverse samples.

\section{Discussion and implications}

Overall, this study provides valuable insights for both practitioners and researchers, highlighting the importance of code comments and their potential to improve the helpfulness and usability of code snippets in online programming communities.

From a research perspective, as motivated in Section~\ref{sec:summary-related-work}, the impact of different ways of presenting and documenting the usually short code snippets on Stack Overflow has not yet been adequately studied.
Findings from larger open-source projects do not necessarily generalize to the specific context of shorter and isolated snippets.
For example, \cite{DBLP:journals/tosem/HuangGDSCLZZ23} found no difference between the perceived helpfulness of inline and block comments, whereas we found that block comments were perceived as more helpful.

Since users' assessment of the perceived helpfulness of code snippets can influence their decision to reuse them, our results have implications beyond Stack Overflow.
Our study points to the context-dependence of perceived helpfulness and code comprehension, highlighting the need for further studies to understand how users decide to reuse code---from Stack Overflow answers or from outputs of AI-based assistants offering alternative solutions. 

From a practical perspective, the findings suggest that developers and contributors on platforms such as Stack Overflow should prioritize adding code comments, with a particular emphasis on block comments.
These types of comments were perceived by users as more helpful, indicating their potential to increase the overall helpfulness and understandability of code snippets.
The guidelines provided by Stack Overflow on writing good answers\footnote{\url{https://stackoverflow.com/help/how-to-answer}} do not mention code comments at all (as of July 2025).
Our suggestion is to extend these instructions based on our findings.

Our study has implications beyond code reading and comprehension.
With the widespread adoption of generative AI tools in software development~\citep{gen-ai-software-development}, the question arises of how to prompt these tools to generate the most helpful solutions matching users' needs.
An interesting question is whether prompting generative AI tools to include block comments generally leads to the generation of more helpful code snippets than not prompting them to include these comments.

The study presented in this article lays the groundwork for different future research directions.
We have investigated the impact of comment types and surface features on the perceived helpfulness of Stack Overflow posts.
However, other features such as comment length, variable naming conventions, or other coding style elements, may also influence perceived helpfulness.
Experiments similar to ours might help to investigate the respective features.
Furthermore, replications of our study, as well as additional qualitative studies, can help refine our results and their interpretation.

On a meta-level, with the advance of generative AI in software engineering, a solid theoretical embedding of activities such as reading and comprehending code becomes even more important for software engineering research.
Without solid theories~\citep{lorey2022theoriesse} of how users carry out central software engineering tasks, it will be difficult to compare the performance of AI-based tools and human developers and to develop suitable benchmarks.

\section{Conclusion}

The goal of this study was to investigate the factors that influence how Stack Overflow users assess the helpfulness of code snippets and solutions. To this end, we developed a simulated Stack Overflow environment that closely resembled the platform's layout and functionality.
The primary focus was on the role of source code comments---specifically how different comment types (block, inline, or none) affect perceived helpfulness. In addition, we examined whether surface-level features on the platform---namely answer position and score---affect users' helpfulness assessment.
We conducted an online experiment with 91 participants using this simulated environment. The results showed that code snippets with block comments were perceived as more helpful than those with inline comments, and both types were rated as more helpful than uncommented code. In contrast, answer order and scores had no significant effect on perceived helpfulness.

Our findings underscore the importance of code comments in influencing how developers judge the helpfulness of code snippets. These insights have practical implications for improving user guidance on platforms like Stack Overflow and, as motivated above, for enhancing the design of AI-based coding assistants that generate or recommend code snippets.

Directions for future research include conducting new experiments that incorporate additional features that may influence perceived helpfulness, as well as evaluating how prompting GenAI tools to include line or block comments impacts the perceived helpfulness of generated code snippets.
Furthermore, replications and additional qualitative studies can help refine our results, working towards detailed theories of how users read, evaluate, and reuse code snippets.


\section{Declarations}

\paragraph{Funding:}
This research received no specific grant from any funding agency in the public, commercial, or not-for-profit sectors.

\paragraph{Ethical Approval:}
This study did not require formal approval from an ethics committee. The research was conducted in accordance with the principles of good scientific practice.

\paragraph{Informed Consent:}
All participants were informed about the purpose of the study and provided their consent prior to participation through the online questionnaire.

\paragraph{Consent to Participate:}
All participants gave informed consent before taking part in the study via the online experiment interface.

\paragraph{Author Contributions:}
\begin{description}
    \item[\bf Kathrin Figl:] Conceptualization, Data curation, Formal analysis, Investigation, Resources, Visualization, Writing (original draft), Writing (review \& editing).
    \item[\bf Maria Kirchner:] Conceptualization, Data curation, Formal analysis, Investigation, Methodology, Resources, Visualization, Writing (original draft), Writing (review \& editing).
    \item[\bf Sebastian Baltes:] Conceptualization, Writing (original draft), Writing (review \& editing).
    \item[\bf Michael Felderer:] Conceptualization, Writing (review \& editing).
\end{description}

\paragraph{Conflict of Interest:}
The authors declare that Sebastian Baltes is a member of the Empirical Software Engineering editorial board.

\paragraph{Clinical Trial Number:}
Not applicable.

\paragraph{Data Availability Statement}
The questionnaire used in our online experiment, along with the raw data analyzed using SPSS, is available as supplementary material~\citep{figl_2024_13319937}.
To protect participant anonymity, we modified any rows that could reveal their identities, including the removal of exact job descriptions and the generalization of participants’ exact ages into age ranges.


\bibliographystyle{spbasic}      
\bibliography{literature-r2.bib}   

\end{document}